# TONGUE-PLACED TACTILE BIOFEEDBACK SUPPRESSES THE DELETERIOUS EFFECTS OF MUSCLE FATIGUE ON JOINT POSITION SENSE AT THE ANKLE

Nicolas VUILLERME, Matthieu BOISGONTIER, Olivier CHENU, Jacques DEMONGEOT

and Yohan PAYAN

Laboratoire TIMC-IMAG, UMR UJF CNRS 5525, La Tronche, France

Address for correspondence:

Nicolas VUILLERME

Laboratoire TIMC-IMAG, UMR UJF CNRS 5525, Faculté de Médecine, 38706 La Tronche

cédex, France.

Fax: (33) (0) 4 76 51 86 67

Email: nicolas.vuillerme@imag.fr

**Abstract**

Whereas the acuity of the position sense at the ankle can be disturbed by muscle fatigue, it recently also has been shown to be improved, under normal ankle neuromuscular state, through the use of an artificial tongue-placed tactile biofeedback. The underlying principle of this biofeedback consisted of supplying individuals with supplementary information about the position of their matching ankle position relative to their reference ankle position through electrotactile stimulation of the tongue. Within this context, the purpose of the present experiment was to investigate whether this biofeedback could mitigate the deleterious effect of muscle fatigue on joint position sense at the ankle. To address this objective, sixteen young healthy university students were asked to perform an active ankle-matching task in two conditions of No-fatigue and Fatigue of the ankle muscles and two conditions of No-biofeedback and Biofeedback. Measures of the overall accuracy and the variability of the positioning were determined using the absolute error and the variable error, respectively. Results showed that the availability of the biofeedback allowed the subjects to suppress the deleterious effects of muscle fatigue on joint position sense at the ankle. In the context of sensory re-weighting process, these findings suggested that the central nervous system was able to integrate and increase the relative contribution of the artificial tongue-placed tactile biofeedback to compensate for a proprioceptive degradation at the ankle.

3**Introduction**

Proprioceptive afferent from the ankle joint is known to play a significant role in the control of human posture and gait that greatly influence our ability to perform activities of daily living. Impaired ankle proprioception also may be a predisposing factor for chronic ankle instability, balance difficulties, reduced mobility functions, fall, injury and re-injury (e.g. Fu et al. 2005; Halasi et al. 2005; Madhavan and Shields 2005; Payne et al. 1997). Interestingly, recent studies reporting deleterious effects of localized fatigue of the ankle muscles on joint position sense (Forestier et al. 2002) and balance control (e.g. Ledin et al. 2004; Vuillerme et al. 2006a; Vuillerme and Demetz 2007; Vuillerme et al. 2002a,b, 2003) support this view. Accordingly, there is a need to develop techniques for enhancing proprioceptive acuity at the ankle. A promising intervention, using an artificial tongue-placed tactile biofeedback, has recently been shown to improve ankle joint position sense in young healthy adults (Vuillerme et al. 2006b,c). The underlying principle of this system consisted of supplying individuals with supplementary information about the position of their matching ankle position relative to their reference ankle position through a tongue-placed tactile output device generating electrotactile stimulation of the tongue (Tongue Display Unit, TDU) (Bach-y-Rita et al. 1998; Bach-y-Rita and Kercel 2003) (Figure 1). Note that the tongue was chosen as a substrate for electrotactile stimulation site according to its neurophysiologic characteristics. Indeed, because of its dense mechanoreceptive innervations (Trulsson and Essick, 1997) and large somatosensory cortical representation (Picard and Olivier, 1983), the tongue can convey higher-resolution information than the skin can (Sampaio et al. 2001; van Boven and Johnson 1994). In addition, due to the excellent conductivity offered by the saliva, electrotactile stimulation of the tongue can be applied with much lower voltage and current than is required for the skin (Bach-y-Rita et al. 1998). At this point however, although the above-mentioned studies evidenced that this artificial tongue-placed tactile biofeedback can be used to improve ankle joint proprioception under normal ankle neuromuscular state (i.e. with redundant and reliable sensory information) (Vuillerme et al. 2006b,c), its effectiveness under altered ankle proprioceptive conditions, as it is the case following muscle fatigue (Forestier et al. 2002), has not been established yet. Within this context, the purpose of the present experiment was to investigate whether this biofeedback could mitigate the deleterious effect of muscle fatigue on joint position sense at the ankle. It was hypothesized that (1) ankle muscles fatigue would decrease ankle joint position sense without the provision of the biofeedback delivered through the TDU, (2) biofeedback would improve ankle joint position sense in the absence of ankle muscles fatigue, and, more originally (3) biofeedback would mitigate the detrimental effect of ankle muscles fatigue on ankle joint position sense.

**Methods**

Subjects

Sixteen male young healthy university students (age = 22.3 ± 2.4 years; body weight = 72.1 ± 9.5 kg; height = 180.8 ± 6.2 cm) voluntarily participated in the experiment. Criteria for selection and inclusion were: male; aged 20-30 years; negative medical history; normal ankle range of motion. Exclusion criteria were: history of sprain, fracture, injury, surgery of the lower extremities; restricted ankle range of motion of less than 20° of dorsiflexion or 30° of plantarflexion; history of central or peripheral nervous system dysfunctions. Subjects were familiarized with the experimental procedure and apparatus, and they gave their informed consent to the experimental procedure as required by the Helsinki declaration (1964) and the local Ethics Committee. None of the subjects presented any history of injury, surgery or pathology to either lower extremity that could affect their ability to perform the ankle joint position sense test.

Apparatus for measuring ankle joint position sense

Ankle joint position sense measurements were carried out as previously done by Vuillerme et al. (2006b,c). Subjects were seated barefoot in a chair with their right and left foot secured to two rotating footplates with Velcro straps. The knee joints were flexed at about 110°. Movement was restricted to the ankle in the sagittal plane, with no movement occurring at the hip or the knee. The axes of rotation of the footplates were aligned with the axes of rotation of the ankles. Precision linear potentiometers attached on both footplates provided analog voltage signals proportional to the ankles' angles. A press-button was held in the right hand and served to record the matching. Signals from the potentiometers and the press-button were sampled at 100 Hz (12 bit A/D conversion), then processed and stored within the Labview 5.1 data acquisition system. In addition, a panel was placed above the subject's legs to eliminate visual feedback about both ankles position.

Apparatus for providing tactile biofeedback

The underlying principle of the biofeedback device, similar to that recently used by Vuillerme et al. (2006b,c), consisted of supplying individuals with supplementary information about the position of the matching right ankle relative to the reference left ankle position through a tongue-placed tactile output device. Electrotactile stimuli were delivered to the front part of the tongue dorsum via flexible electrode arrays placed in the mouth, with connection to the stimulator apparatus via a flat cable passing out of the mouth. The system comprised a 2D electrode array ($1.5 \times 1.5$ cm) consisting of 36 gold-plated contacts each with a 1.4 mm diameter, arranged in a $6 \times 6$ matrix (Vuillerme et al. 2006b,c, 2007a,b,c) (Figure 1).

-----------------------------------

Insert Figure 1 about here

-----------------------------------

The following coding scheme for the TDU was used (Vuillerme et al. 2006c): (1) no electrical stimulation when both ankles were in a similar angular position within a range of 0.5°; (2) stimulation of either the anterior or posterior zone of the matrix (2 × 6 electrodes) (i.e. stimulation of front and rear portions of the tongue) depending on whether the matching right ankle was in a too plantarflexed or dorsiflexed position relative to the reference left ankle, respectively. The frequency of the stimulation was maintained constant at 50 Hz across subjects, ensuring the sensation of a continuous stimulation over the tongue surface. Conversely, the intensity of the electrical stimulating current was adjusted for each subject, and for each of the front and rear portions of the tongue, given that the sensitivity to the electrotactile stimulation was reported to vary between individuals (Essick et al. 2003), but also as a function of location on the tongue in a preliminary experiment (Vuillerme et al. 2006b).

Task and procedure

The experimenter first placed the left reference ankle at a predetermined angle where the position of the foot was placed on a wooden support. By doing so, subjects did not exert any effort to maintain the position of the left reference ankle, preventing the contribution of effort cues coming from the reference ankle to the sense of position during the test (e.g., Vuillerme et al. 2006b,c; Walsh et al. 2004; Winter et al. 2005). A matching angular target position of 10° of plantarflexion was selected to avoid the extremes of the ankle range of motion to minimize additional sensory input from joint and cutaneous receptors (e.g., Burke et al. 1988). Once the left foot had been positioned at this test angle, subjects' task was to match its position by voluntary placement of their right ankle. When they felt that they had reached the target angular position (i.e., when the right foot was presumably aligned with the left foot), they were asked to press the button held in their right hand, thereby registering the matched position. Subjects were not given feedback about their performance and were not given any speed constraints other than 10-s delay to perform one trial. Results of a preliminary study using 10 subjects established the "excellent" test-retest reliability of the active ankle-matching task (intra-class correlation coefficient (ICC) > 0.75).

Two sensory conditions were presented, (1) the No-biofeedback serving as a control condition and (2) the Biofeedback condition in which the subjects executed the active matching task using a TDU-biofeedback system, as described above. These two sensory conditions also were executed under two states of ankle muscles fatigue. (1) The No-fatigue condition served as a control condition, whereas (2) in the Fatigue condition, the measurements were performed immediately after a fatiguing procedure. Its aim was to induce a muscular fatigue in the ankle plantar-flexor of the right leg until maximal exhaustion. Subjects were asked to perform toe-lifts with their right leg as many times as possible following the beat of a metronome (40 beats/min). Verbal encouragement was given to ensure that subjects worked maximally. The fatigue level was reached when subjects were no more able to complete the exercise. Immediately on the cessation of exercise, the subjective exertion level was assessed through the Borg CR-10 scale (Borg 1990). Subjects rated their perceived fatigue in the ankle muscles as almost "extremely strong" (mean Borg ratings of 9.1 ± 0.6). To ensure that ankle joint proprioception measurements in the Fatigue condition were obtained in a real fatigued state, various rules were respected, as described in previous studies investigating the effect of ankle muscles fatigue on postural control (Ledin et al. 2004; Vuillerme et al. 2006a; Vuillerme and Demetz 2007; Vuillerme et al. 2003). (1) The fatiguing exercise took place beside the experimental set-up to minimise the time between the exercise-induced fatiguing activity and the proprioceptive measurements, (2) the duration of the postfatigue conditions was short (1 minute) and (3) the fatiguing exercise was repeated prior to each No-biofeedback and Biofeedback condition.

For each condition of No-biofeedback and Biofeedback and each condition of No-fatigue and Fatigue, subjects performed five trials, for a total of 20 trials. The order of presentation of the two No-biofeedback and Biofeedback conditions was randomised over subjects.

Data analysis

Two dependent variables were used to assess matching performances (Schmidt 1988).

(1) The absolute error (AE), the absolute value of the difference between the position of the right matching ankle and the position of the left reference ankle, is a measure of the overall accuracy of positioning.

(2) The variable error (VE), the variance around the mean constant error score, is a measure of the variability of the positioning. Decreased AE and VE scores indicate increased accuracy and consistency of the positioning, respectively (Schmidt, 1988).

Statistical analysis

Data obtained for AE and VE were submitted to separate 2 Biofeedback (No-biofeedback *vs.* Biofeedback) × 2 Fatigues (No-fatigue *vs.* Fatigue) analyses of variances (ANOVAs) with repeated measures of both factors. Post hoc analyses (Newman-Keuls) were performed whenever necessary. Level of significance was set at 0.05.

**Results**

Positioning accuracy

Analysis of the AE showed a significant interaction of Biofeedback × Fatigue ($F(1,15) = 4.72$, $P < 0.05$, Fig. 2A). The decomposition of the interaction into its simple main effects showed that the availability of the biofeedback suppressed the effect of Fatigue (Fig. 2A). The ANOVA also showed main effects of Biofeedback ($F(1,15) = 32.65$, $P < 0.001$) and Fatigue ($F(1,15) = 23.89$, $P < 0.001$), yielding smaller AE in the Biofeedback than No-biofeedback condition and larger AE in the Fatigue than No-fatigue condition, respectively.

Positioning variability

Analysis of the VE showed a significant interaction of Biofeedback × Fatigue ($F(1,15) = 6.41$, $P < 0.05$, Fig. 2B). The decomposition of the interaction into its simple main effects showed that the availability of the biofeedback suppressed the effect of Fatigue (Fig. 2B). The ANOVA also showed main effects of Biofeedback ($F(1,15) = 12.70$, $P < 0.01$) and Fatigue ($F(1,15) = 12.60$, $P < 0.01$), yielding smaller VE in the Biofeedback than No-biofeedback condition and larger VE in the Fatigue than No-fatigue condition, respectively.

-----------------------------------

Insert Figure 2 about here

-----------------------------------

**Discussion**

Whereas the acuity of the position sense at the ankle can be disturbed by muscle fatigue (Forestier et al. 2002), it recently also has been shown to be improved, under normal ankle neuromuscular state, through the use of an artificial tongue-placed tactile biofeedback (Vuillerme et al. 2006b,c). The underlying principle of this biofeedback consisted of supplying individuals with supplementary information about the position of their matching ankle position relative to their reference ankle position through electrotactile stimulation of the tongue.

Within this context, the purpose of the present experiment was to investigate whether this biofeedback could mitigate the deleterious effect of muscle fatigue on joint position sense at the ankle.

To address this objective, sixteen young healthy university students were asked to perform an active ankle-matching task in two conditions of No-fatigue and Fatigue of the ankle muscles and two conditions of No-biofeedback and Biofeedback. Measures of the overall accuracy and the variability of the positioning were determined using the absolute error and the variable error, respectively (Schmidt 1988).

On the one hand, without the provision of biofeedback (No-biofeedback condition), fatigue yielded less accurate and less consistent matching performances, as indicated by increased absolute (Fig. 2A) and variable errors (Fig. 2B) observed in the Fatigue relative to the No-fatigue condition, respectively. This result, in line with that of Forestier et al. (2002), was expected (*hypothesis 1*) and also is consistent with the existing literature reporting impaired proprioceptive acuity following muscle fatigue induced at the knee (Lattanzio et al. 1997; Skinner et al. 1986), hip (Taimela et al. 1999), neck (Vuillerme et al. 2005b), shoulder (Björklund et al. 2000; Lee et al. 2003; Voight et al. 1996), or elbow (Allen and Proske 2005; Brockett et al. 1997; Walsh et al. 2004). At this point, although the exact mechanism inducing proprioceptive impairment subsequent to the fatiguing exercise is rather difficult to answer and was not within the scope of our study, it is likely that the above-mentioned adverse effects of muscle fatigue on joint position sense stem from alterations of signals from muscle spindles (Björklund et al. 2000; Brockett et al. 1997; Forestier et al. 2002; Vuillerme et al. 2005b) and/or sense of effort (Allen and Proske 2006; Walsh et al. 2004) associated with muscle fatigue.

On the other hand, in the absence of fatigue at the ankle joint (No-fatigue condition), more accurate and more consistent matching performances were observed with than without biofeedback, as indicated by decreased absolute (Fig. 2A) and variable errors (Fig. 2B), respectively. This result also was expected (*hypothesis 2*), corroborating those of Vuillerme et al. (2006b,c). It confirms that, under "normal" proprioceptive conditions, the central nervous system is able to integrate an artificial biofeedback signal delivered through electrotactile stimulation of the tongue to improve joint position sense at the ankle.

Finally, and more originally, the availability of the biofeedback allowed the subjects to suppress the deleterious effects of muscle fatigue on ankle positioning accuracy and variability, as indicated by the significant interactions Biofeedback × Fatigue observed for absolute (Fig. 2A) and variable errors (Fig. 2B), respectively. Confirming our *hypothesis 3*, these results suggest an increased reliance on biofeedback delivered through the TDU for the active ankle-matching task in condition of muscle fatigue at the ankle. We could interpret these findings relative to the concept of "sensory re-weighting" (e.g. Horak and Macpherson 1996; Oie et al. 2002; Vuillerme et al., 2002a,2005a). Indeed, it has been proposed that, as a sensory input becomes lost or disrupted (as it was the case for the proprioceptive signals from the ankle in the Fatigue condition), the relative contributions of alternative available sensory inputs are adaptively re-weighted by increasing reliance on sensory modalities providing accurate and reliable information (as it was the case when the tactile biofeedback delivered through the TDU was made available in the Biofeedback condition). Interestingly, although the experimental task was different, this interpretation is reminiscent with previous results evidencing the adaptive capabilities of the central nervous system to cope with ankle muscle fatigue for controlling posture during bipedal quiet standing. It was observed that, to some extent, vision (Ledin et al. 2004; Vuillerme et al. 2006a), ankle foot orthoses (Vuillerme and Demetz 2007) and a light finger touch (Vuillerme and Nougier 2003) could compensate for the postural perturbation induced by muscular fatigue applied to the ankle, hence reflecting a re-weighting of sensory cues in balance control following ankle muscles fatigue by increasing the reliance on visual information, cutaneous inputs from the foot and shank and haptic cues from the finger, respectively. From a fundamental perspective in neuroscience, the above-mentioned results and ours collectively argue in favour of a re-weighting of available, accurate and reliable sensory cues as a function of the neuromuscular constraints acting on the individual. At this point, however, while the present experiment was specifically designed to assess the combined effects of muscle fatigue and tongue-placed tactile biofeedback on joint position sense at the ankle, it could also be interesting to investigate these combined effects on the control of ankle movements in more functional tasks such as postural control during quiet standing or gait initiation. Finally, the present findings could also have implications in clinical conditions and rehabilitation practice. Biofeedback delivered through TDU may indeed enable individuals to overcome ankle proprioceptive impairment caused by trauma (Bressel et al. 2004; Halasi et al. 2005; Payne et al. 1997), normal aging (Madhavan and Shields 2005; Verschueren et al. 2002), or disease (Simoneau et al. 1996; van den Bosch et al. 1995; van Deursen et al. 1998). An investigation involving older healthy adults is being conducted to strengthen the potential clinical value of the approach reported in the present experiment.

**Acknowledgements**

The authors are indebted to Professor Paul Bach-y-Rita for introducing us to the Tongue Display Unit and for discussions about sensory substitution. Paul has been for us more than a partner or a supervisor: he was a master inspiring numerous new fields of research in many domains of neuroscience, biomedical engineering and physical rehabilitation. The authors would like to thank subject volunteers. Special thanks also are extended to Damien Flammarion and Sylvain Maubleu for technical assistance, Benjamin Bouvier for his help in data collection, Dr. Vince and Zora B. for various contributions. This research was supported by the Fondation Garches and the company IDS.

**Figure captions**

**Figure 1.** Photograph of the Tongue Display Unit used in the present experiment. It comprises a 2D electrode array (1.5 × 1.5 cm) consisting of 36 gold-plated contacts each with a 1.4 mm diameter, arranged in a 6 × 6 matrix.

**Figure 2**. Mean and standard deviation for the absolute error (A) and the variable error (B) for the two conditions of No-fatigue and Fatigue and the two conditions of No-biofeedback and Biofeedback. The significant *P* values for comparison between No-fatigue and Fatigue conditions are reported (NS: $P > 0.05$; **: $P < 0.01$; ***: $P < 0.001$).

**Figure 1.**

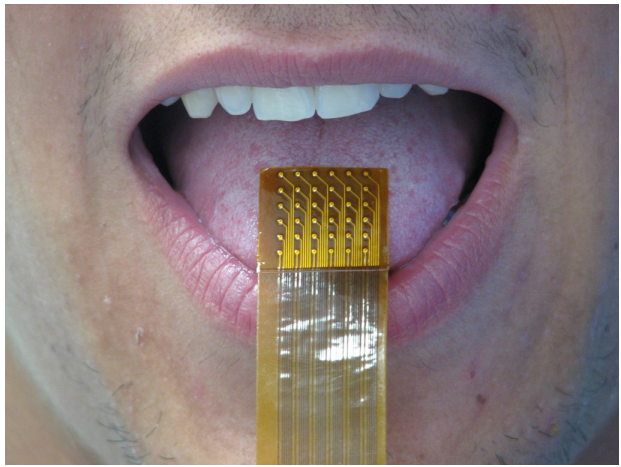

**Figure 2.**

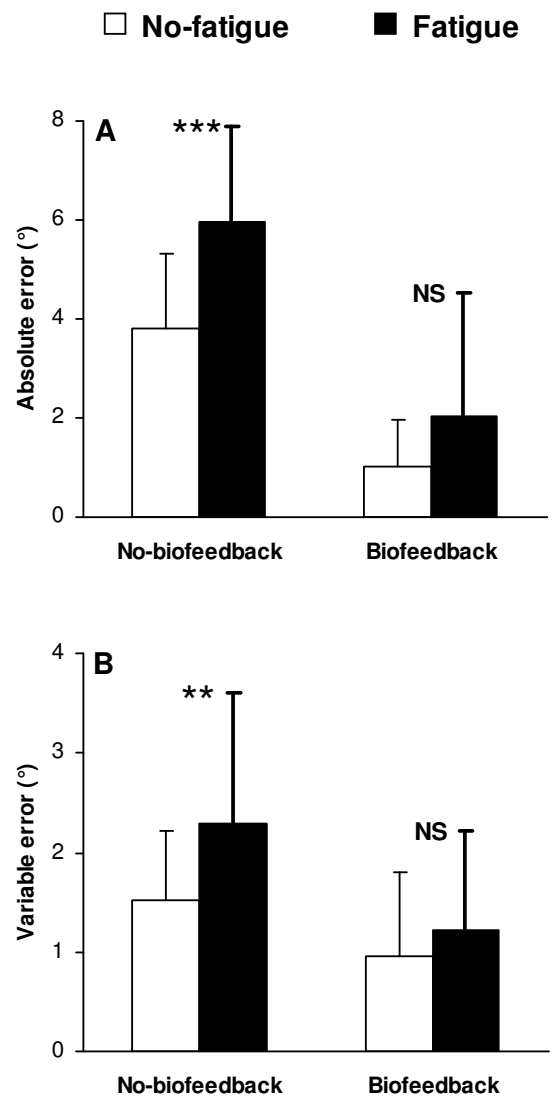